\documentclass[sigconf]{acmart}

\AtBeginDocument{%
  \providecommand\BibTeX{{%
    \normalfont B\kern-0.5em{\scshape i\kern-0.25em b}\kern-0.8em\TeX}}}

\copyrightyear{2021}
\acmYear{2021}
\setcopyright{acmcopyright}
\acmConference[SIGIR '21] {Proceedings of the 44th International ACM SIGIR Conference on Research and Development in Information Retrieval}{July 11--15, 2021}{Virtual Event, Canada.}
\acmBooktitle{Proceedings of the 44th International ACM SIGIR Conference on Research and Development in Information Retrieval (SIGIR '21), July 11--15, 2021, Virtual Event, Canada}
\acmPrice{15.00}
\acmISBN{978-1-4503-8037-9/21/07}
\acmDOI{10.1145/XXXXXX.XXXXXX}

\usepackage{url}
\usepackage{amsthm}
\usepackage{booktabs}
\usepackage{algorithm}
\usepackage{algorithmic}
\usepackage{threeparttable}
\usepackage{tabularx, booktabs}
\usepackage{caption}
\usepackage{subfigure}

\newcommand{\mb}{\mathbf}

\newcolumntype{P}[1]{>{\centering\arraybackslash}p{#1}}
\begin{document}
\fancyhead{}
\title{AdsGNN: Behavior-Graph Augmented Relevance Modeling in Sponsored Search}


\author{Chaozhuo Li}
\affiliation{%
	\institution{Microsoft Research Asia}
	\country{}}
\email{cli@microsoft.com}

\author{Bochen Pang}
\authornote{* Indicates Equal Contributions.}
\affiliation{%
	\institution{Microsoft}
	\country{}}
\email{bochen.pang@microsoft.com}

\author{Yuming Liu}
\email{yumliu@microsoft.com}
\author{Hao Sun}
\email{hasun@microsoft.com}
\affiliation{%
  \institution{Microsoft}
  \country{}}

\author{Zheng Liu}
\email{zheng.liu@microsoft.com}
\author{Xing Xie}
\email{xingx@microsoft.com}
\affiliation{%
	\institution{Microsoft Research Asia}
	\country{}}

\author{Tianqi Yang}
\email{tianqi.yang@microsoft.com}
\author{Yanling Cui}
\email{yanling.cui@microsoft.com}
\affiliation{%
	\institution{Microsoft}
	\country{}}

\author{Liangjie Zhang}
\email{liazha@microsoft.com}
\author{Qi Zhang}
\email{zhang.qi@microsoft.com}
\affiliation{%
	\institution{Microsoft}
	\country{}}


\begin{abstract}
Sponsored search ads appear next to search results when people look for products and services on search engines. 
In recent years, they have become one of the most lucrative channels for marketing. 
As the fundamental basis of search ads, relevance modeling has attracted increasing attention due to the significant research challenges and tremendous practical value.  
Most existing approaches solely rely on the semantic information in the input query-ad pair, while the pure semantic information in the short ads data is not sufficient to fully identify user's search intents. 
Our motivation lies in incorporating the tremendous amount of unsupervised user behavior data from the historical search logs as the complementary graph to facilitate relevance modeling. 
In this paper, we extensively investigate how to naturally fuse the semantic textual information with the user behavior graph, and further propose three novel AdsGNN models to aggregate topological neighborhood from the perspectives of nodes, edges and tokens.   
Furthermore, two critical but rarely investigated problems, domain-specific pre-training and long-tail ads matching, are studied thoroughly.   
Empirically, we evaluate the AdsGNN models over the large industry dataset, and the experimental results of online/offline tests consistently demonstrate the superiority of our proposal. 
\end{abstract}


\keywords{sponsored search; relevance modeling; graph mining}

\maketitle

\section{Introduction}

Search engine advertising has become a significant element of
the web browsing due to the outburst of search demands. 
Choosing the right ads for a query and the order in which they are displayed greatly affects the probability that a user will see and click on each ad \cite{ling2017model}. 
As the fundamental component of sponsored search systems, the relevance model measures the semantic closeness between an input query and a candidate ad, which is capable of improving the user experience and driving revenue for the advertisers.

Existing relevance models usually capture the semantic correlations inside the query-ad pairs with powerful Natural Language Understanding (NLU) models, which has been a key research field with many breakthroughs over the last years. 
Deep Structured Semantic Model (DSSM) \cite{shen2014learning, liang2018dynamic} is one of the first powerful solutions to encode text data into latent distributed vectors. 
With the development of NLU, transformers \cite{vaswani2017attention} and BERT \cite{devlin2018bert,lu2020twinbert} are both the emerging pre-trained language models with far superior performance, which surpass previous approaches and can even approach the human level. 
Pre-trained models can capture the contextual information in the sentences and generate high-quality language representations, leading to the promising performance.

However, directly applying such NLU approaches to the relevance modeling scenario may not be desirable as the pure semantic information in the ads data is not sufficient to fully identify the user search intents. 
The queries and ads are quite short (e.g., less than 4 words on average from Bing's log) compared with the long text (e.g., sentences and documents) used in the traditional NLU tasks, and thus the performance is hindered by the scarce semantic information in the  input short texts. 
Manually labeled query-ad pairs can provide more linguistic guidance for better understanding the latent search intents, but they are quite expensive and time-consuming. 
Besides, although existing models have demonstrated satisfying results in matching common queries with popular ads, they usually achieve undesirable performance on the long-tail queries/ads  \cite{zhu2021textgnn}, which is potentially caused by under-training due to naturally scarce data on these low-frequency examples.    

Existing NLU-based relevance models mainly focus on implicit feature engineering solely from the input textual data.  
Structural complexities of new relevance models grow exponentially but the performance improvement is relatively marginal. 
Therefore, in order to significantly  improve the relevance performance, employing new and accessible data with supplementary information is a more practical and favorable approach. 
A natural and easily accessible data source that provides information beyond semantic text in the search engine system is users' historical click behaviors. 
Our motivation lies in exploring this cheap and massive click data as complementary for relevance modeling. 
For example, given a short query ``$AF1$'', it is intractable for NLU models to understand the actual meanings correctly. 
But in the user behavior data, its historical clicked ads include ``$Air \ Force \ 1$'' and ``$Nike \ AF1\ shoes$''.
With this complementary information, the input query can be easily comprehended as a sub-brand of Nike shoes, which will facilitate the downstream relevance modeling task.

A straightforward strategy is to employ click relations as a surrogate of relevance annotations.
Namely, the clicked query-ad pair is viewed as positive, and negative pairs are synthesized by fixing a query while randomly sampling its un-clicked ads. 
This strategy confuses the relevance correlations with the click relations and thus may introduce ambiguities from two aspects. 
Firstly, the arbitrariness and subjectivity of user behavior lead to the misalignment between user clicks and true relevance annotations \cite{li2019learning}, which may introduce noises into the ground truth and further pollute the training set. 
Secondly, negative pairs sampled by data synthesizing usually share no common tokens for queries and ads, which may mislead the relevance model to view common terms as critical evidence of relevance. However, lexically similar query and ad may have totally different intents such as ``iPad'' and ``iPad charger''.

Here we extensively study how to naturally incorporate the user behaviors in the format of graphs without introducing the mentioned ambiguities.  
Queries and ads are connected by click relations, based on which a bipartite behavior graph can be built easily from the search logs.  
Over the recent years, Graph Neural Network (GNN) \cite{velivckovic2017graph, hamilton2017inductive} and the variants are widely applied on graph structural data with promising performance on many downstream applications (e.g., node and graph classifications).   
Inspired by GNN, we attempt to properly integrate the abundant behavior graph data into the traditional semantic model.  
In most GNN models, the node textual features are pre-existed and fixed in the training phase. 
In contrast, we make the semantic and graph models work in conjunction with each other, which contributes to generating more comprehensive representations for deeper intent understanding.

In this paper, we propose the AdsGNN models that naturally extend the powerful text understanding model with the complementary graph information from user historical behaviors which serves as a powerful guide to help us better understanding the latent search intents. 
Firstly, three variations are proposed to effectively fuse the semantic textual information and user behavior graph from the perspectives of nodes, edges and tokens. 
Specifically, the node-level AdsGNN$_{n}$ views the queries and ads as separated entities, then aggregates the neighborhood information from the click graph to facilitate the learning of entity representations. 
Edge-level AdsGNN$_{e}$ treats the input query-ad pair as a unit and directly learns the edge representations by aggregating the local topological information.  
Token-level AdsGNN$_{t}$ regards the input token as processing units and fuses the graph information to learn behavior-enhanced token representations, which is a deeper tightly-coupled integration approach compared with previous two variations. 
After that, in order to take full advantage of user behaviors, we propose to pre-train the AdsGNN model on the behavior graph with two graph-based pre-training objectives. 
Finally, as the long-tail queries/ads are usually associated with limited click signals, we propose to learn a knowledge distillation model to directly learn the mapping from the pure semantic space to the behavior-enhanced embedding space, which empowers our proposal with powerful topological predictive ability to handle the long-tail entities.         
Empirically, AdsGNNs are evaluated over the large industry dataset and yield gains over state-of-the-art  baselines with regards to the different parameter settings, which demonstrates the superiority of our proposal.

We summarize our main contributions as follows.
\begin{itemize}
	\item 
	We incorporate the unsupervised user behavior graph as complementary to enrich the semantic information in the input query-ad pairs and propose three novel AdsGNN models to effectively fuse textual data and behavior graph from different perspectives.     
	
	\item  A novel knowledge distillation based framework is proposed to alleviate the formidable but less explored challenge of long-tail ad matching.     
	
	\item  Extensively, we evaluate our proposal on the large industry dataset. Experimental results demonstrate the superior performance of the proposed AdsGNN models.
\end{itemize}

\section{Problem Definition} 
In this section, we will formally define the studied problem. 
Different from existing approaches solely based on the semantic data, here we further exploit the unsupervised user behavior data from the search log as complementary.
The behavior graph is defined as a bipartite graph: $\mathbf{G} = \{\mathbf{Q}, \mathbf{A}, \textbf{E}\}$, in which  $\mathbf{Q}$ and $\mathbf{A}$ denote the set of queries and ads, respectively. $\textbf{E} \in \mathbb{N} ^{|\mathbf{Q}| \times |\mathbf{A}|}$ is the adjacency matrix, which includes the click signals between queries and ads. 
The ground truth of relevance modeling task is defined as a set of triples: $\mathbf{L} =\{<q_{i}, a_{i}, y_{i}> \}$, where $q_{i}$ denotes the input query, $a_{i}$ denotes the ad and $y_{i} \in \{0,1\}$ represents the relevance label between $q_{i}$ and $a_{i}$. 
We aim to learn a classifier $\mathit{f}$: $\mathit{f}(q,a) \in \{0,1\}$ by fusing the ground truth  $\mathbf{L}$ and the click graph $\mathbf{G}$.

\section{Methodology}

\begin{figure}
	\centering
	\subfigure[The illustration of AdsGNN$_{n}$ in the behavior graph.]{
		\label{fig:nodelvel} 
		\includegraphics[width=0.3\textwidth]{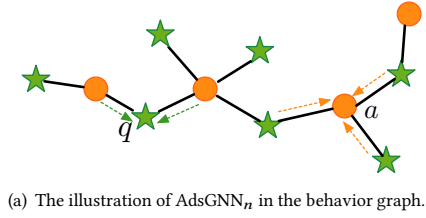}}
	\hspace{0.01in}
	\subfigure[Overview of the model structure.]{
		\label{fig:nodemodel} 
		\includegraphics[width=0.47\textwidth]{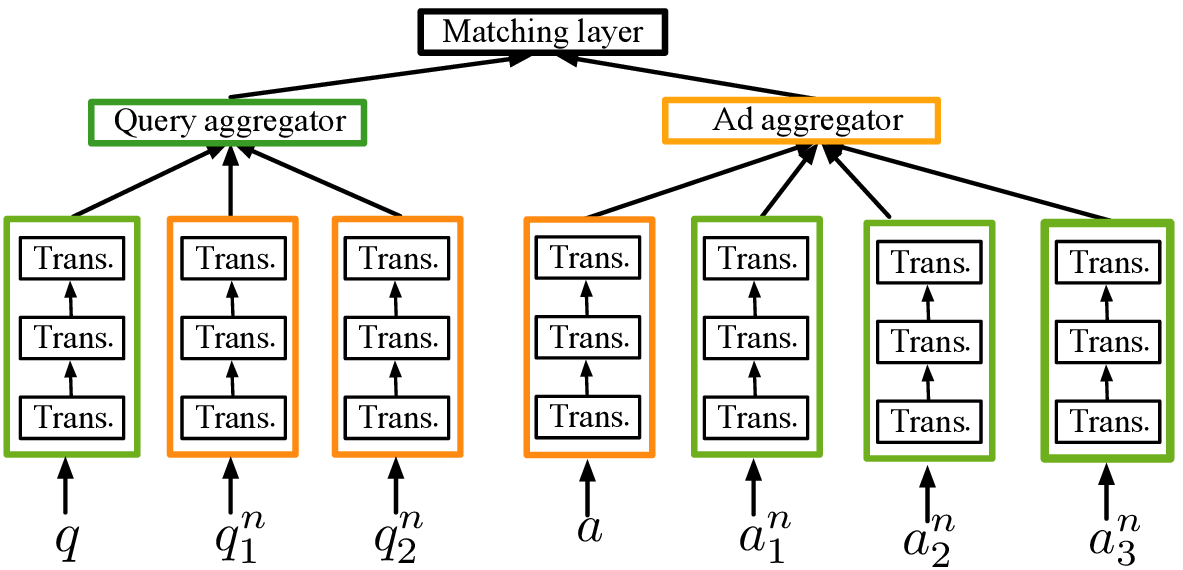}}
	\caption{The illustration of the node-level AdsGNN$_{n}$ model.}
	\label{fig:node} 
\end{figure}

\subsection{AdsGNN$_{n}$ for Node-level Aggregation} 

Denote $q$ and $a$ as the input query and ad, respectively. 
Figure \ref{fig:nodelvel} shows the motivation of the node-level aggregation model AdsGNN$_{n}$. 
Given the input $q$ (green pentagram), its neighbors in the behavior graph are the ads (orange circles) that users tend to click.
The dotted lines represent message-passing directions. 
Considering the scarce semantic information in the short input text, fusing the behavior neighborhood contributes to correctly understanding the user's search intents.

The architecture of the AdsGNN$_{n}$ model is illustrated in Figure \ref{fig:nodemodel}. 
$q_{i}^{n}$ is the $i$-th graph neighbor (ad) of the input query $q$, and $a_{i}^{n}$ is the $i$-th graph neighbor (query) of the input ad $a$.  
The proposed model is a natural extension of the high performance two-tower models (e.g., C-DSSM \cite{gao2015deep} and TwinBERT \cite{lu2020twinbert}) with additional information from graph structural data. 
As there exist millions of candidate ads, it is infeasible to use a sophisticated text encoder to compute the similarity between a search query and each ad one-by-one  \cite{lu2020twinbert}. 
Hence, the twin-tower structure is a good choice for online serving as we could pre-compute the ad representations in advance. When a query comes, we can easily generate its embedding and calculate the similarities between the input query embedding and cached ad representations.      
From the bottom up, AdsGNN$_{n}$ model includes the following three layers: 

\noindent \textbf{Node Encoder.} 
Node encoder embeds the textual information of input entities into the low-dimensional latent vectors, which can be implemented as any layer-wise text encoding models.   
For example, as shown in Figure \ref{fig:nodemodel}, BERT is selected as the node encoder. 
The input text is first tokenized by the WordPiece \citep{wu2016google} tokenizer. 
For each token within the input sequence, the initial embedding is acquired with the summation of its token embedding and positional embedding before it goes through three transformer encoder layers. 
After the node encoder layer, we can get a sequence of vectors corresponding to the tokens in the sentence. 
The vectors are then combined using a weighted-average pooling layer following TwinBERT model. 
Considering the topology context of queries and ads are quite different (e.g., the number of a query's neighbors is usually larger than the one of an ad), we design two node encoders with different parameter sets marked with different colors.  Encoders with the same color are set to share the same parameters.

\noindent \textbf{Node Aggregator.}
The output vectors of the node encoders are fed into the node aggregator to perform neighborhood fusion. 
A desirable node aggregator should be able to collect valuable information from the behavior neighborhood, then fuse these messages with the center node to learn high-quality contextual representations. 
Considering the importance of different neighbors differ greatly in depicting the local context,  we employ the self-attention strategy used in Graph Attention Network (GAT) \cite{velivckovic2017graph} to learn the neighbor weights properly and then weighted combine the contextual semantic information.  
The attention score between the input query $q$ and its neighbor $q_{i}^{n}$ is calculated as follows: 
\begin{equation}
\alpha_{qi} = \frac{\exp (\sigma(\mb{a}_{q} \cdot [\mb{h}_{q}||\mb{h}_{qi}]))}{\sum_{i = 1}^{N} \exp(\sigma(\mb{a}_q \cdot [\mb{h}_{q}||\mb{h}_{qi}]))}
\end{equation}
in which $||$ is the concatenation  operation, $\mb{a}_{q}$ is the local-level attention vector for the neighbors and $\sigma$ is activation function to introduce nonlinear transformation. $N$ denotes the number of neighbors. $\mb{h}_{q}$ and $\mb{h}_{qi}$ are the embeddings learned by node encoder of $q$ and $q_{i}^{n}$ , respectively. 
The learned attention score $\alpha_{qi}$ denotes how important neighbor $q_{i}^{n}$ will be for the center node $q$.
Then the contextual embedding $\mb{z}_{q}^{n}$ can be achieved by aggregating the neighbor’s encoded features with the corresponding
coefficients:
\begin{equation}
\mb{z}_{q}^{n} = \sigma(\sum_{i = 1}^{N} \alpha_{qi}\cdot \mb{h}_{qi})  
\end{equation}
This contextual embedding is then connected with the direct output
of the query encoder through the concatenation, similar to the idea of Residual Connection Network \cite{he2016deep}:
\begin{equation}
\begin{aligned}
\mb{z}_{q} = [\mb{h}_{q}||\mb{z}_{q}^{n} ]
\end{aligned}
\end{equation} 
Note that, query and ad aggregators are implemented with different parameter sets following the bipartite nature of the behavior graph.

\noindent \textbf{Matching Layer.} The topology-augmented query representation is interacted with ad-side output in the matching layer to get the final classification outputs. Here we implement the matching layer as a multi-layer perceptron (MLP) with one-hidden layer following previous works \cite{lu2020twinbert,zhu2021textgnn,li2016detecting}.  

AdsGNN$_{n}$ first learns the semantic embedding of each input text, and then combine them together following the guidance of user historical behaviors. 
The aggregation is processed in the node-level in the GNN style, which is expected to provide additional information beyond the traditional semantic texts.  

\subsection{AdsGNN$_{e}$ for Edge-level Aggregation} 

\begin{figure}
	\centering
	\subfigure[The illustration of AdsGNN$_{e}$ in the behavior graph.]{
		\hspace{0.4in}
		\label{fig:edgelvel} 
		\includegraphics[width=0.35\textwidth]{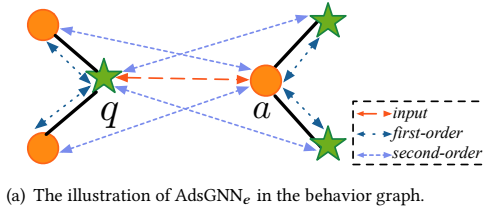}}
	\hspace{0.01in}
	\subfigure[Overview of the model structure.]{
		\label{fig:edgemodel} 
		\includegraphics[width=0.45\textwidth]{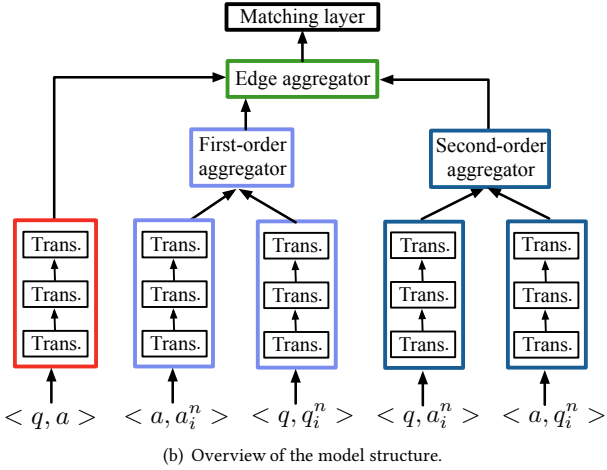}}
	\caption{The illustration of the edge-level AdsGNN$_{e}$ model.}
	\label{fig:edge} 
\end{figure}

The proposed AdsGNN$_{n}$ model aggregates the neighborhood information on the node-level, in which input queries and ads are processed separately. 
However, the studied relevance matching problem focuses on the correlations inside the query-ad pair instead of a single query or ad. 
Handling the query-ad pair as a whole can capture and exploit the information from both sides, while the node-based model can only utilize the data from one side.    
Query-ad pairs can be naturally viewed as the edges in the behavior graph, and thus we propose  AdsGNN$_{e}$  for the edge-level aggregation.    

In order to embed the topological context into edges as much as possible, we design the following three types of edges as shown in Figure \ref{fig:edgelvel}:  
\begin{itemize}
	\item \textbf{Input Edge} denotes the relevance relation between the input query $q$ and ad $a$, and most existing works are solely based on this connection. 
	
	\item \textbf{First-order Edge} captures the click relations between the center node and its neighbors in the behavior graph. 
	The insight lies in the homophily \cite{thelwall2009homophily,li2019adversarial} among the connected edges.    
	Given the input query $q$, its neighbor $q_{i}^n$ and the input ad $a$ can be viewed as the co-clicked items, and thus these two adjacent edges <$q, a$> and <$q, q_{i}^n$> tend to share similar search intents. 
	Thus, the first-order edge can capture the local context of the input edge as the complementary.

	\item \textbf{Second-order Edge} encodes the semantic relations between the input query (ad) and the neighbors of the input ad (query) (e.g., <$q, a_{i}^n$> and <$a, q_{i}^n$>). Our motivation is that if two ads are both clicked given the same query, they have a larger chance to be similar. Second-order edges are capable of capturing the semantic information from high-order neighbors to enrich the inputs. 
\end{itemize}

Different meaningful and complex semantic information is involved in the various types of edges, and different edges may also extract diverse semantic information. 
It is nontrivial to select the most meaningful edges and fuse the semantic information suitably.  
Hence, we propose a novel AdsGNN$_{e}$ model based on the hierarchical attention, including first/second-order and edge-level attentions. 
Specifically, the first/second-order attentions aim to learn the importance of edges in the same category, while the edge-level attention is able to learn the informativeness of different types of edges and assign proper weights to them.
Figure \ref{fig:edgemodel} demonstrates the details of  AdsGNN$_{e}$ model, which includes the following four layers:

\noindent \textbf{Edge Encoder.} 
Edge encoder aims to capture the textual information of the input edge along with the correlations between the associated query and ad.  
The text of query and ad in the input edge are concatenated with the [SEP] token to define the sentence boundary as $[query, SEP, ad]$. 
After processing with the tokenizer, a three-layer BERT is employed to learn the representation of the input edge. 
In the transformer layer, each token on one side (e.g., query) is able to attend tokens on the other side (e.g., ad), which captures the correlations inside query-ad pair and is intrinsically different from the AdsGNN$_{n}$ model.    
Note that, we design different encoders for different types of edges to emphasize the various roles.  

\noindent \textbf{First/Second-order Aggregator.} 
The output vectors of the edge encoders are fed into the first/second-order aggregators based on the edge categories. 
The first-order aggregator is selected as an example, which is designed to learn the importance of different first-order edges. 
Denote the embedding of the $i$-th first-order edge as $\mb{h}_{ei}$. 
Different from the node-level aggregator, these edges are  position-insensitive without the definitions of center or neighbor edges.    
Hence, we design another format of attention as follows: 
\begin{equation}
\label{word_attention}
\begin{aligned}
a_{ei} &= \tanh(\mb{w}_{e} \cdot \mb{h}_{ei} + b_{e}), \\
\alpha_{ei} &= \frac{\exp(a_{ei})}{\sum_{j=1}^{N}\exp(a_{ej})} \\ 
\mb{s}_{e} &= \sum_{i=1}^{N} \alpha_{ei} \mb{h}_{ei}
\end{aligned}
\end{equation}
in which $N$ is the number of first-order edges, $\mb{w}_{e}$ and $b_{e}$ are the trainable parameters.  
$\alpha_{ei}$ represents the normalized importance of the $i$-th edge compared with other edges. 
Edges belonging to the same category are aggregated according to the  attention scores. 

\noindent \textbf{Edge Aggregator.} 
Different types of edges may have different significance, and thus we design the edge aggregator to distinguish informative edge types from less informative ones.  
Here we also employ the self-attention method to learn the weights of three edge types, which is similar to the attention mechanism used in the node aggregator. 
The input edge is viewed as the center edge and others are regards as the neighbors.     

Overall, AdsGNN$_{e}$ model can be understood as the natural extension of the traditional one-tower models (e.g., BERT), which usually achieve higher performance by considering the rich information from both sides but cannot be easily applied on the low-latency online matching scenario. 

\begin{figure}
	\centering
	\includegraphics[width=0.47\textwidth]{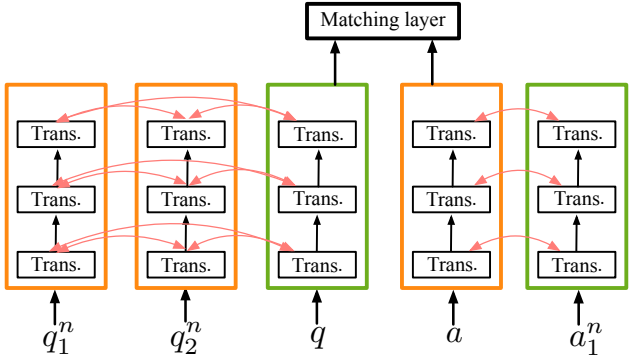}
	\caption{The illustration of token-level AdsGNN$_{t}$ model.}
	\label{fig:tokenlevel} 
\end{figure}

\subsection{AdsGNN$_{t}$ for Token-level Aggregation}

The previous two models first learn the embeddings of input entities based on their own textual data and then aggregate the neighbor embeddings as the final representations. 
Graph aggregation module is built on top of the textual encoder, which forms a cascade framework. 
In each layer of the encoder, a token can only attend to other tokens in the belonged node but cannot refer to the tokens in other nodes, leading to a loosely-coupled structure. 
Here we aim to deeply fuse the textual and topological information with a tightly-coupled structure named AdsGNN$_{t}$, where multiple layers of graph transformers and text transformers are alternately deployed.  

\begin{figure}
	\centering
	\includegraphics[width=0.45\textwidth]{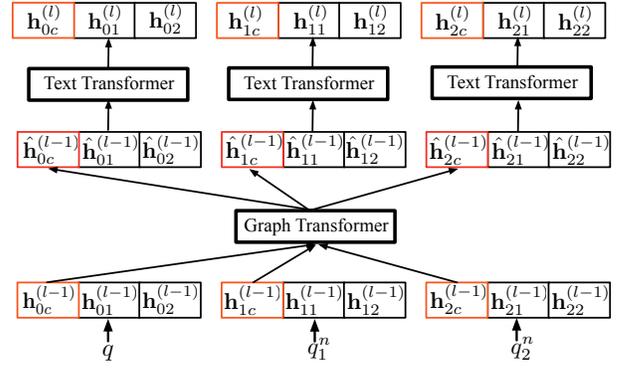}
	\caption{The details of the $l$-th AdsGNN$_{t}$ layer.}
	\label{fig:tokenmodel} 
\end{figure}

As shown in Figure \ref{fig:tokenlevel}, messages are passed among contextual neighbors and the center node in each node encoder layer. 
Hence, tokens within the center node can attend to tokens in the neighborhood, which contributes to learning the topology-aware token representations. 
Each input text is tokenized into a sequence of tokens by WordPiece \citep{wu2016google}. 
For each tokenized sequence, a [CLS] token is padded in the front. 
The embedding of [CLS] token is regarded as the representation of the input text. 
As shown in Figure \ref{fig:tokenmodel}, the $i$-th layer of AdsGNN$_{t}$ on the query side is selected as an example, which includes the following two components:

\noindent \textbf{Graph Transformer.}   
Graph transformer aggregates the contextual topology  information, and then dispatches it to the input texts.  
Firstly, we define the symbol $\mb{h}_{ij}^{(l)}$ shown in Figure \ref{fig:tokenmodel}, which represents the embedding of $j$-th token in the $i$-th node in the layer $l$. Index $i$ is set to $0$ for the center node and $j=c$ means this is the embedding of CLS token.    
The CLS embeddings $\mb{h}_{ic}^{(l-1)}$ from the ($l$-1)-th layer can be organized into a matrix $\mb{H_{c}}^{(l-1)} = [\mb{h}_{0c}^{(l-1)}, \mb{h}_{1c}^{(l-1)}, \cdots, \mb{h}_{Nc}^{(l-1)}]^\top \in \mathbb{R}^{(N+1) \times d_h}$, in which $N$ is the number of neighbors and $d_{h}$ denotes the dimension of latent embedding. 
Then, the graph-transformer is introduced below to exchange information among different nodes: 
\begin{equation}
\begin{aligned}
\hspace{-5pt} \hat{\mb{H}}_{c}^{(l-1)} &= \mbox{G-Transformer} \left( \mb{H}_{c}^{(l-1)}\right)\\
&= \mbox{softmax} \left(\frac{\mb{Q} \mb{K}^\top}{\sqrt{d_h}} \right) \mb{V},
\end{aligned}
\end{equation}
where
\begin{equation}
\begin{cases}
\mb{Q} &= \mb{H}_{c}^{(l-1)} \mb{W}_Q^{(l-1)},\\
\mb{K} &= \mb{H}_{c}^{(l-1)} \mb{W}_K^{(l-1)},\\
\mb{V} &= \mb{H}_{c}^{(l-1)} \mb{W}_V^{(l-1)}.\\
\end{cases}
\end{equation}
In the above equations, $\mb{W}_Q^{(l-1)}, \mb{W}_K^{(l-1)}, \mb{W}_V^{(l-1)} \in \mathbb{R}^{d_h \times d_h}$ denote the trainable variables. 
The learned matrix $\hat{\mb{H}}_{c}^{(l-1)}$ preserves the topological context information. 
Finally, the contextual token embedding vector $\hat{\mb{h}}_{ic}^{(l-1)}$ is padded with the embedding sequence of the rest tokens as shown in the middle part of Figure \ref{fig:tokenmodel}. 

\noindent \textbf{Text Transformer.}     
The contextual CLS embedding $\mb{\hat{h}}_{ic}^{(l-1)}$ captures the semantics of tokens in other nodes. 
Text transformer aims to transmit this contextul information from the CLS token to the textual ones within an input sentence. 
Here we employ the vanilla transformer model as the text transformer. 
The self-attention mechanism is able to covey the contextual information from the padded CLS embedding to the rest tokens. 
The CLS embedding of the input query/ad in the last layer is outputted as the final representation and then fed into the matching layer to fit the ground truth dataset.  

Overall, AdsGNN$_{t}$ utilizes the embedding of a special token as the intermediate to convey messages among tokens in different nodes. 
In the text transformer phase, the CLS token collects the information from other textual tokens in the belonged sentence, and then exchange them with CLS tokens in other nodes by graph transformer. 
This indirect message passing structure is designed to ensure model efficiency. 
A straight-forward approach is to directly concatenate all the tokens from the neighborhood together and feed them into the BERT model. 
Denote $m$ as the average length of the input text and $n$ as the number of neighbors. 
Each token will attend to $(m \times (n+1))$ tokens, which significantly enlarges the parameter space and will be seriously memory-cost and time-consuming. 
However, the attended field sizes of graph transformer and text transformer are $n$ and $m$, respectively, which ensures our proposal is capable of both aggregating neighbor tokens and maintaining model efficiency.  

\subsection{Objective Function} 
The output vector from the matching layer is denoted as $\mb{y}'\in \mathbb{R}^{1 \times 2} $, which contains the predicted  probabilities of the input pair is relevant or not. 
We select the cross-entropy as the objective function:
\begin{equation} \label{eq:3}
\begin{split}
\mathcal{L} &= \sum_{x \in \mb{L}} cross(\mb{y}, \mb{y}'), \\
cross (\mb{y}, \mb{y}') &= - \sum_{i} \mb{y}_{i}\log(\mb{y}_{i}').
\end{split}
\end{equation}

\section{Pre-training on Behavior graph}
The text encoders used in our proposal are set to the transformers,  so we can conveniently load the weights from the first three layers
of the pre-trained large BERT model to get a good starting point. 
However, the available BERT checkpoints are pre-trained on the general text dataset, which is intrinsically different from the sponsored search scenario. 
The ad text is much shorter than the normal sentences, and the relevance relation between the query and ad is also different from the adjacency relationship captured by the popular Next Sentence Prediction (NSP) task \cite{devlin2018bert}. 
Hence, we propose the following two objective functions to adapt the general pre-trained checkpoints to the unsupervised user behavior graph as the domain-specific warm up.  

The first task is the neighbor-enhanced masked language modeling (NE-MLM), which aims to exploit the topological neighbors to predict the masked tokens. 
We first randomly select a set of entities from the click graph as the center node and then collect their neighbor texts. 
After that, a subset of the center node's textual tokens is replaced by the mask token. 
The BERT model is expected to predict the masked tokens according to the remaining ones and the data from neighborhood.   
NE-MLM empowers the text encoders with the ability to capture the token-level topological connections.  

The second task is the neighbor prediction (NP), which is similar to the NSP task and only differs in the input data. 
Our motivation lies in predicting the neighbors given the center node, which captures the topological relations at the entity-level. 
We first randomly sample a set of edges from the behavior graph. 
Text in a node is viewed as the next sentence of the text in its connected node and vice visa. 
The constructed text pairs are then used to strengthen the model's ability to leverage the rich user behaviors. 
After pre-training the BERT checkpoint with these two tasks, we select the first three layers as the initialization of the text transformers used in the AdsGNN model.   

\begin{figure}
	\centering
	\subfigure[The illustration of the teacher model.]{
		\label{fig:teacher} 
		\includegraphics[width=0.30\textwidth]{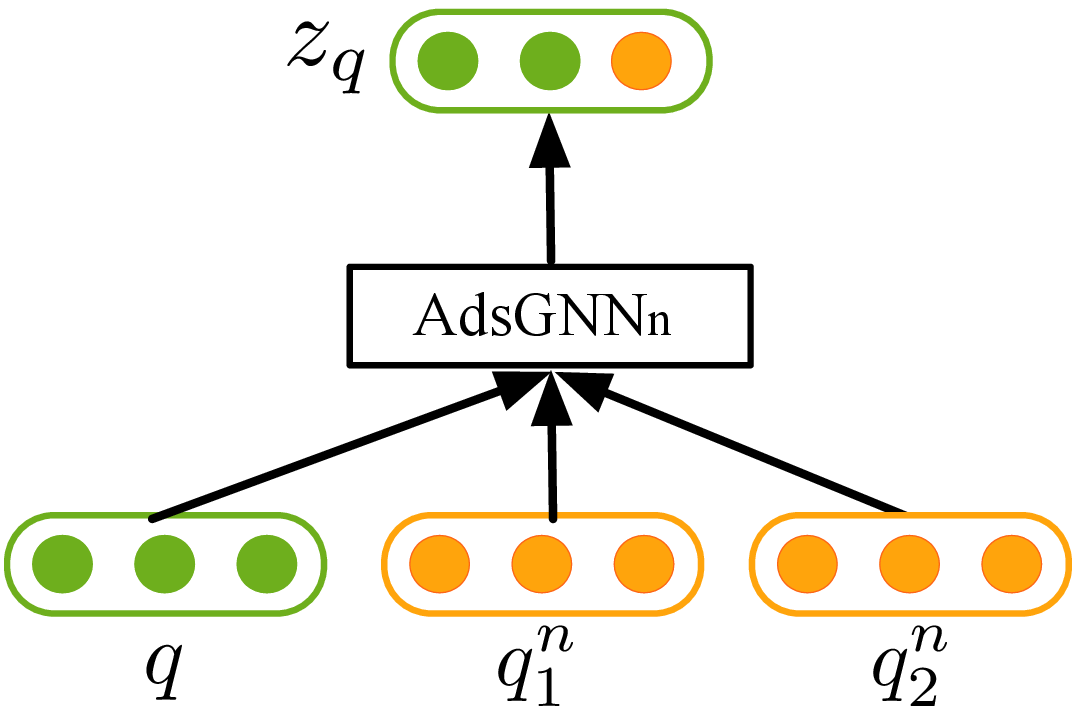}}
	\hspace{0.01in}
	\subfigure[Student model.]{
		\label{fig:student} 
		\includegraphics[width=0.136\textwidth]{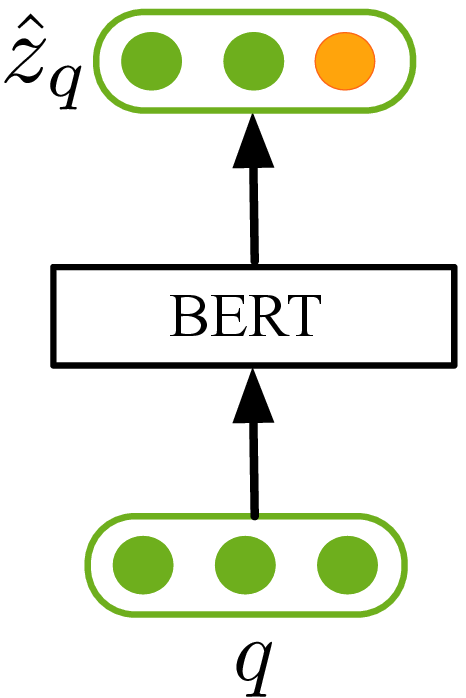}}
	\caption{The illustration of  knowledge distillation model.}
	\label{fig:kd} 
\end{figure}

\section{Knowledge Distillation for Long-tail Ads Matching}

The proposed AdsGNN models leverage the unsupervised user behavior graph to enhance the relevance matching performance. 
However, their practical value is potentially hindered by the low-frequency long-tail entities. 
Neighborhood information from the behavior graph is indispensable to our proposals in both the training and inference phases, while long-tail entities are usually associated with few or even no neighbors. 
This scarce topological information brings a formidable challenge to the graph-enhanced semantic models, which is rarely studied by previous works.      

Assume the AdsGNN$_{n}$ model has been fully trained on the normal ads dataset. 
Next, we will study how to apply the learned model to generate quality embeddings for long-tail entities in the inference phase. 
One straight-forward approach is to set the missing neighbors as a special token (e.g., [PAD]) and feed them into the learned model to obtain the representations. 
Given an input query $q$ and its neighbors $q_{i}^{n}$, AdsGNN$_n$ essentially learns the mapping function $\mathcal{M}$: $\mb{z}_{q} = \mathcal{M}(q, q_{i}^{n})$, in which $\mb{z}_{q}$ is the learned embedding and will be fed into the matching layer. 
If we input the query $q$ with meaningless tokens into the learned mapping function to get $\mb{z}'_{q} = \mathcal{M}(q)$, it is obvious that $\mb{z}_{q}$ and $\mb{z}'_{q}$ would be quite different, which may further lead to the diverse matching results. 
Thus, this padding strategy may not be a desirable solution. 
Another possible solution is to comprehend the long-tail entities with their semantical  closeness neighbors. 
For example, we can use Approximate Nearest Neighbor (ANN) search \cite{fu2016efanna} to select the top similar ads $\tilde{q}_{i}^{n}$ from the whole candidate set and view them as the complemented neighbors of $q$.  
Despite of the low-efficiency of ANN search on the dataset with millions of samples, the mapping function $\mathcal{M}$ is trained to capture the click relations between the queries and ads, which are intrinsically  different from the semantic similarities. 
Thus, the set $\tilde{q}_{i}^{n}$ and $q_{i}^{n}$ have a large chance to be different, leading to the undesirable matching performance.

Here we aim to alleviate the challenge of long-tail ads matching in the manner of knowledge distillation \cite{hinton2015distilling}. 
As shown in Figure \ref{fig:teacher},  a pre-trained  AdsGNN$_{n}$ is viewed as the teacher model. 
The learned representation $\mb{z}_{q}$ encodes the input query and its neighbors, which will be used as the training target of the student model.       
The right sub-figure illustrates the student model, which is a vanilla BERT model with the output as $\hat{\mb{z}}_{q}$. 
Our motivation lies in that the output vector $\hat{\mb{z}}_{q}$ of the student model should be similar to $\mb{z}_{q}$ as much as possible. 
The objective function is defined as follows: 
\begin{equation}
\begin{aligned}
\mathcal{L}_{kd} = ||\hat{\mb{z}}_{q} - \mb{z}_{q}|| ^{2}
\end{aligned}
\end{equation} 
The L2-loss $\mathcal{L}_{kd}$ forces the outputs of the student model should be topological preserving, which strengthens the student model with the ability to latently predict the neighbor information. 
Thus, it can effectively handle the long-tail entities. 
In addition, according to the universal approximation theorem \cite{hornik1991approximation}, a feed-forward neural network can approximate any nonlinear functions and BERT is a comparatively sophisticated model with a large number of parameters. Hence, the student model is expected to  approximate the mapping function learned from the teacher model successfully.

\section{Experiments}

\begin{table}
	\centering
	\begin{threeparttable}
		\caption{Statistics of the ground truth dataset. }
		\begin{tabular}{cccc}
			\toprule
			\multicolumn{1}{c}{}&\multicolumn{1}{c}{Postive samples}&\multicolumn{1}{c}{Negative samples}&\multicolumn{1}{c}{All}\cr
			\midrule
			Training &  184,213 & 410,021 & 594,234 \\
			Validation &  21,798 & 44,237 & 66,026 \\
			Test &  59,028 & 127,770 & 186,798 \\
			\bottomrule
		\end{tabular}
		\label{tab:statistics_of_datasets}
	\end{threeparttable}
\end{table}

\subsection{Dataset} 

The proposed AdsGNN models are extensively evaluated on a large industry dataset and 
Table \ref{tab:statistics_of_datasets} shows its statistics. 
This dataset is imbalanced, which is consistent with the fact that most query-ad pairs are unrelated.  
User behavior graph is sampled from the search log of a commercial search engine with \textbf{300 million} distinct segments and more than \textbf{1 billion} clicked relations.   
There exist some publicly available datasets like MSLR\footnote{https://www.microsoft.com/en-us/research/project/mslr/} and ESR\footnote{https://data.world/crowdflower/ecommerce-search-relevance}. 
However, they only contain the relevance annotations but lack the historical user behaviors, which are not suitable for the studied graph-augmented task.    
Although only one dataset is adopted, it includes \textbf{847,058} different manually labeled samples, which is much larger than the publicly available datasets (e.g., 32,000 for ESR).  
Besides, it is an online serving relevance dataset used by a popular commercial search engine and achieves satisfying online performance, which proves the quality and comprehensiveness of this dataset.

\subsection{Baseline Methods}
We select several  state-of-the-art supervised baselines, including the semantic-based, graph-based and hybrid models. 
The semantic-based baselines are shown as follows: 
\begin{itemize}
	\item \textbf{C-DSSM} \cite{shen2014latent} is a latent semantic model that
	incorporates a convolutional-pooling structure over word
	sequences to learn representations for queries and documents. 
	
	\item \textbf{BERT} \cite{devlin2018bert} is a dominant method of pre-training language representations. Here we concatenate the query and ad as the input of BERT model.  
	
	\item \textbf{Twin-BERT} \cite{lu2020twinbert} is two-tower BERT-based structure model, which serves for the efficient retrieval. 
\end{itemize} 
Our proposal is also compared with the graph neural networks. 
Following the normal GNN pipeline, we first learn the semantic embeddings of entities with the pre-trained BERT checkpoint, which are viewed as static node features. 
Then, GNN models aggregate the contextual node features into the center one.    
The objective function is to maximize the similarities between query and ad in the positive pairs while minimizing the similarity scores in negative pairs. 
Two hybrid approaches are also introduced as follows: 
\begin{itemize}
	\item \textbf{MetaBERT}  concatenates the text of center node and contextual neighbors as the input of a BERT model. 
	
	
	\item \textbf{TextGNN} \cite{zhu2021textgnn} incorporates the text and graph information with a node-level aggregator, which is similar to AdsGNN$_{n}$. The major difference is that query and ad encoders in TextGNN share the same parameters, while  AdsGNN$_{n}$ views them as independent modules.    
\end{itemize}	

For the proposed AdsGNN model, ``Bert-base-uncased'' in the huggingface\footnote{https://github.com/huggingface/transformers/} is selected as pre-trained BERT model. 
The number of behavior neighbors is set to 3. 
For the pre-training tasks, 15\% of the input tokens are masked (80\% of them are replaced by the mask token, the rest are replaced randomly or kept as the original tokens with the same probabilities). 
The size of minimal training batch is 64, learning rate is set to 0.00001, number of training epochs is set to 3. 
The student model in the knowledge distillation part is implemented with a five-layer BERT model. 
Adam optimizer \cite{kingma2014adam} is employed to minimize the training loss. 
Other parameters are tuned on the validation dataset and we save the checkpoint with the best validation performance as the final model.  
Parameters in baselines are carefully tuned on the validation set to select the most desirable parameter setting.  
Considering the high imbalance distribution of the annotations, following the previous work \cite{li2019learning} we select ROC-AUC score as the measurement, which represents the area under the Receiver Operating Characteristic curve. We release our code to facilitate future research (\url{https://github.com/qwe35/AdsGNN}).

\begin{table}
	\small
	\centering
	\begin{threeparttable}
		\caption{Performance with different training ratio $T_{r}$.}
		\begin{tabular}{c|ccccc}
			\toprule
			$T_{tr}$&$T_{tr}$=0.1&$T_{tr}$=0.3&$T_{tr}$=0.5&$T_{tr}$=0.7&$T_{tr}$=1.0\cr
			\midrule
			GAT&0.706&0.713&0.726&0.736&0.756\cr
			GraphSAGE&0.702&0.715&0.729&0.737&0.751\cr
			GraphBert&0.764&0.783&0.791&0.805&0.811\cr
			\midrule
			C-DSSM&0.762& 0.795 & 0.803 & 0.812 &0.829 \cr
			Bert&0.786&0.816&0.833&0.839&0.858\cr
			Twin-BERT&0.779&0.814&0.827&0.835&0.852\cr
			\midrule
			MetaBERT&0.792&0.823&0.834&0.846&0.862\cr
			TextGNN&0.801&0.831&0.839&0.848&0.857\cr
			\midrule
			AdsGNN$_{n}$&0.815&0.843&0.849&0.855&0.863\cr
			AdsGNN$_{e}$&0.820&0.846&0.857&0.863&0.873\cr
			AdsGNN$_{t}$&\textbf{0.829}&\textbf{0.852}&\textbf{0.864}&\textbf{0.869}&\textbf{0.881}\cr
			\bottomrule
		\end{tabular}
		\label{tab:trainingratio}
	\end{threeparttable}
\end{table}

\subsection{Experimental Results} 
$T_{tr}$ portion of the training set is randomly selected to train the relevance models. 
We repeat this process three times and report the average ROC-AUC scores. 
Training ratio $T_{tr}$ is increased from 0.1 to 1.0. Experimental results are reported in Table \ref{tab:trainingratio} with different settings of $T_{tr}$. 
From the results, one can see that GNN models obtain the worst performance, which may be due to the node textual features are pre-existed and fixed in the training phase, leading to the limited expression capacity.   
The one-tower textual model (BERT) outperforms the two-tower models (C-DSSM and Twin-BERT) as it can incorporate the information from both sides, while two-tower models can only exploit the data from a single side. 
However, one-tower structure has to  compute the similarity between a search query and each ad one-by-one, which is not suitable for low-latency online scenario. 
By integrating the user behavior graph with semantic information under a unified co-training framework, the hybrid models beat the other baselines by nearly 1\%.   
Our proposals consistently outperform all the baselines. 
Compared with TextGNN, the node-level aggregation model AdsGNN$_{n}$ can capture the different roles of queries and ads, leading to around 1\% performance gain. 
AdsGNN$_{e}$ surpasses AdsGNN$_{n}$ by nearly 0.7\% because it is capable of capturing the correlations between queries and ads.   
AdsGNN$_{t}$ achieves the best performance, demonstrating  that the tightly-coupled structure is more powerful than the loosely-coupled framework in deeply fusing the graph and textual information.  
From the results, we can summarize the following two conclusions: (1) the unsupervised user behaviors indeed facilitate the relevance matching task; 
(2) it would be better to integrate the complementary information into the underlying semantic units (e.g., tokens) instead of the high-level ones (e.g., nodes and edges).

\subsection{Ablation Study} 
Here we perform the ablation study on the proposed models from different perspectives. 
The training ratio $T_{r}$ is set to 0.1. 

\begin{table}
	\centering
	\begin{threeparttable}
		\caption{Ablation study on the pre-training tasks. }
		\begin{tabular}{ccccc}
			\toprule
			\multicolumn{1}{c}{}&\multicolumn{1}{c}{AdsGNN$_n$}&\multicolumn{1}{c}{AdsGNN$_e$}&\multicolumn{1}{c}{AdsGNN$_t$}\cr
			\midrule
			None &  0.804 & 0.812 & 0.819 \\
			NE-MLM &   0.812 & 0.818 & 0.825 \\
			NP &   0.807 & 0.815 & 0.823 \\
			NE-MLM + NP &  \textbf{0.815} & \textbf{0.820} & \textbf{0.829} \\
			\bottomrule
		\end{tabular}
		\label{tab:pre-training}
	\end{threeparttable}
\end{table}   

\begin{table}
	\centering
	\begin{threeparttable}
		\caption{Performance of different combinations of edge types. }
		\begin{tabular}{P{4cm}P{4cm}}
			\toprule
			\multicolumn{1}{c}{Edge combination}&\multicolumn{1}{c}{ROC-AUC}\cr
			\midrule
			$e_{in}$ &0.786\cr
			$e_{in}$ + $e_{1st}$&0.817\cr
			$e_{in}$ + $e_{2nd}$&0.804\cr
			$e_{in}$ + $e_{1st}$ + 
			$e_{2nd}$ &\textbf{0.820}\cr
			\bottomrule
		\end{tabular}
		\label{tab:combinations}
	\end{threeparttable}
\end{table} 

\noindent \textbf{Pre-training Tasks} Here we aim to investigate the importance of two proposed pre-training tasks NE-MLM and NP. 
Text encoders are pre-trained with various combinations of  pre-training tasks. 
Table \ref{tab:pre-training} presents the experimental results. 
One can see that the model performance significantly drops without any pre-training tasks, which demonstrates the effectiveness of the domain-specific adaption.    
NE-MLM task outperforms the NP task, proving that the token-level correlations are more critical than the node-level correlations. 
This conclusion is consistent with the superior  performance of token-level aggregation model AdsGNN$_{t}$. 
Mixing these two tasks can sufficiently incorporate the domain-specific data from different levels, thus achieving the best performance.  

\noindent \textbf{Edge Category} 
Here we perform ablation study on the proposed AdsGNN$_{e}$ model from the perspective of edge category selection. 
As mentioned in section 3.2, three types of edges are proposed to capture the local topology structure: input edge $e_{in}$, first-order edge  $e_{1st}$ and second-order edge $e_{2nd}$. 
Here we investigate the performance of different combinations of three types of edges, in which
$e_{in}$ is indispensable as it is the primary learning target. 
Hence, four ablation models can be obtained as shown in Table \ref{tab:combinations} along with the relevance matching performance.  
Without the neighborhood information, model $e_{in}$ performs the worst.   
Model $e_{in} + e_{1st}$ significantly outperforms $e_{in} + e_{2nd}$, which demonstrates that the directly connected neighbors are more important and informativeness than the long-distance neighbors. 
The proposed AdsGNN$_{e}$ model achieves the best performance by enjoying the merits of three types of edges.

\begin{table}
	\centering
	\begin{threeparttable}
		\caption{Ablation study on the aggregation mechanism. }
		\begin{tabular}{P{3cm}P{2cm}P{2cm}}
			\toprule
			\multicolumn{1}{c}{Aggregation strategy}&\multicolumn{1}{c}{AdsGNN$_{n}$}&\multicolumn{1}{c}{AdsGNN$_{e}$}\cr
			\midrule
			Mean-pooling &0.772&0.785\cr
			Max-pooling&0.791&0.783\cr
			Summation & 0.784 & 0.776\cr
			LSTM &0.802&0.804\cr
			Self-attention & \textbf{0.815} &\textbf{0.820}\cr
			\bottomrule
		\end{tabular}
		\label{tab:attention}
	\end{threeparttable}
\end{table} 

\noindent \textbf{Aggregation Mechanism} 
In the AdsGNN$_{n}$ and AdsGNN$_{e}$ models, self-attention is selected as the node/edge aggregator. 
In order to demonstrate the effectiveness of attention-based aggregator, four popular aggregation strategies are selected as the baselines including mean-pooling, max-pooling, summation and LSTM \cite{hamilton2017inductive}.  
Results are shown in Table \ref{tab:attention}. 
Max-pooling outperforms the mean-pooling and summation operations as it can effectively extract the strongest features from the inputs but suffers from serious information missing. 
LSTM can preserve all the input text information and thus obtain higher performance. 
However, LSTM aggregates the neighbors in sequences, which cannot distinguish the informativeness nodes from the less important ones. 
Self-attention  enables the AdsGNN models to learn the  importance of the input nodes/edges and weighted combine them together as the  high-quality representations, leading to more promising performance.

\begin{figure}
	\centering
	\includegraphics[width=0.37\textwidth]{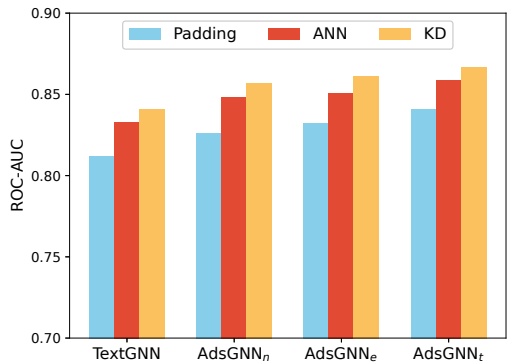} 
	\caption{Results of different completion strategies.}
	\label{fig:longtail} 
\end{figure}

\subsection{Experimental Results on the Long-Tail Set} 
Here we evaluate the performance of the proposed knowledge distillation based model on the long-tail dataset.    
Node degrees in the user behavior graph indicate the clicked frequencies, and thus we randomly select a subset of nodes with only one edge connected as the long-tail entities. 
Then, we combine the selected node with its neighbor as the long-tail query-ad pairs. 
The relevance models are still trained on the normal  training set and then applied to the long-tail dataset. 
Here we design the following three neighborhood completion approaches for the long-tail entities:
\begin{itemize}
	\item \textbf{Padding} adds the special token [PAD] as the neighbors.  
	
	\item \textbf{ANN} selects the most similar entities from the whole set based on the semantic closeness and then add them as the complemented neighbors.
	
	\item \textbf{KD} is the proposed knowledge distillation approach.
\end{itemize}	   

The strongest baseline method TextGNN is selected for comparison. 
Experimental results are shown in Figure \ref{fig:longtail}. 
Given the same neighbor completion strategy, AdsGNN$_{t}$ consistently outperforms other methods, demonstrating that the token-level aggregation is more robust to the scarce topology information.  
ANN method significantly beats the Padding strategy by around 3\%, proving that the semantic-related neighbors can bring richer information than the meaningless special tokens.   
The proposed KD strategy consistently outperforms the other two approaches. 
By viewing the topology-augmented representation as the learning target, the learned KD model is able to fit the mapping function which directly projects the input text from the original textual space to the topology-aware space. 
The neighborhood of  long-tail entities can be latently predicted and incorporated by the KD model.

\begin{figure}
	\centering
	\subfigure[Number of neigbors.]{
		\label{fig:para1} 
		\includegraphics[width=0.225\textwidth]{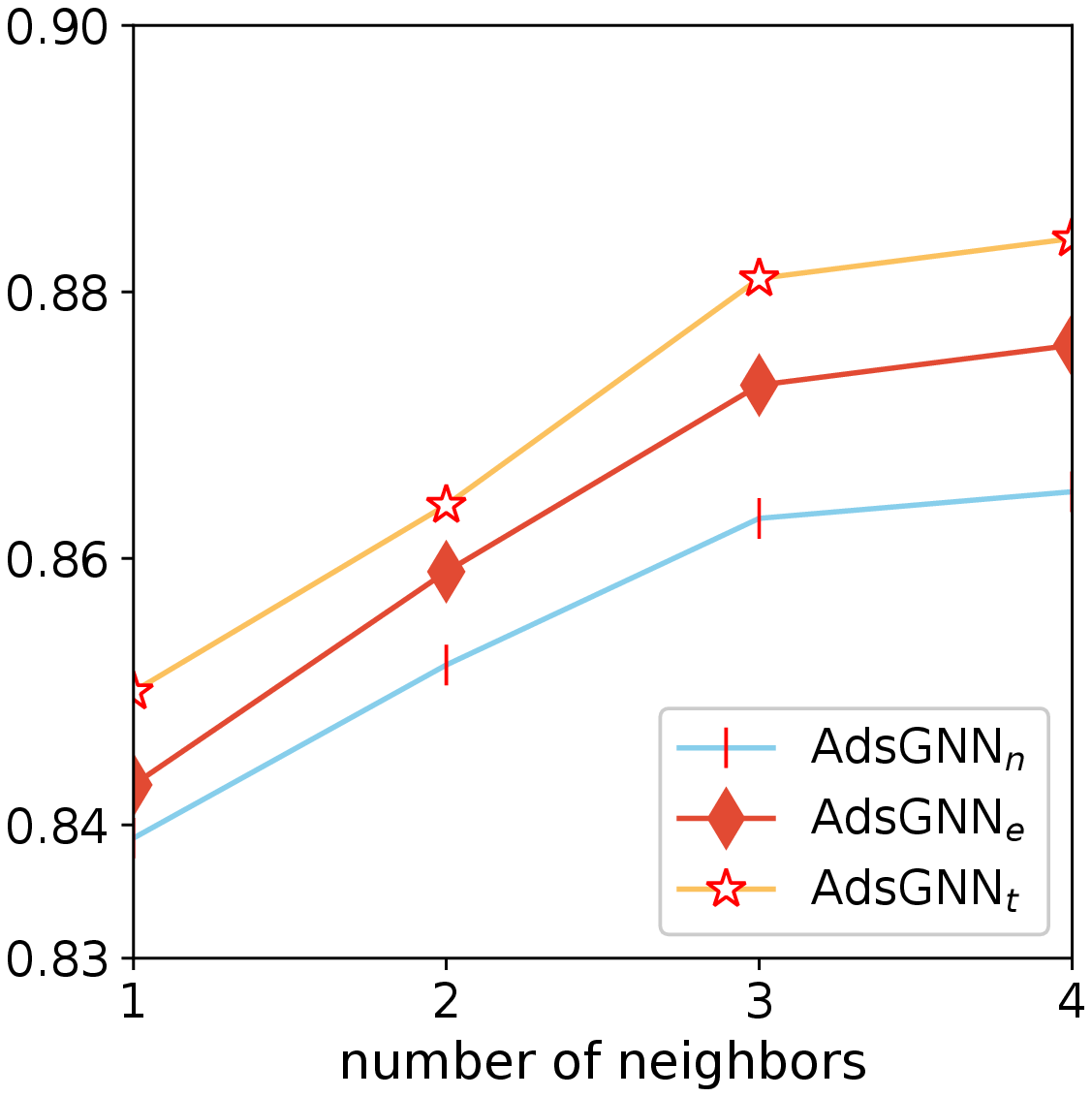}}
	\hspace{0.01in}
	\subfigure[Number of transformer layers.]{
		\label{fig:sparsity_t} 
		\includegraphics[width=0.225\textwidth]{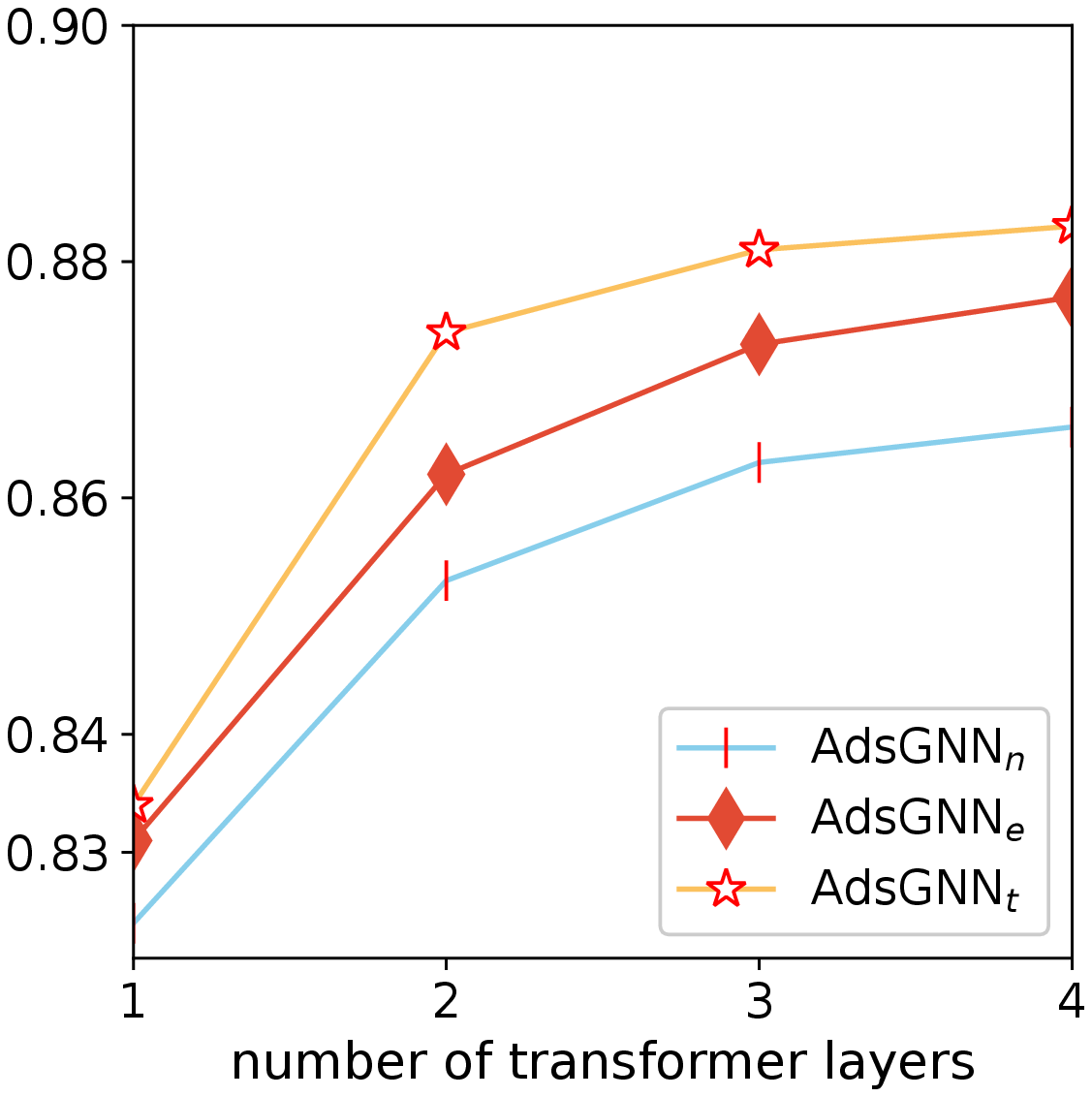}}
	\caption{Parameter sensitivity study.}
	\label{fig:param} 
\end{figure}

\subsection{Parameter Sensitivity Study} 
We study the performance sensitivity of AdsGNN models on two core parameters: the number of neighbors $n$ and the number of transformer layers $l$ used in encoders.  
Training ratio $T_{tr}$ is set to 1. 
Both parameters vary from 1 to 4, and the ROC-AUC scores under different settings  are recorded.  
Figure \ref{fig:param} presents the experimental results. 
From the results, one can see that with the increase of $n$, the performance first increases and then keeps steady. 
It means that the incorporation of more contextual neighbors contributes to better matching performance at the beginning. 
When the available contextual semantic information is fully explored, larger neighbor number cannot further improve model performance, which in turn slows down the training speed. 
When the parameter $l$ increases from 1 to 2, the performance is significantly improved. After that, the trend rate of growth slows down. 

\subsection{Online A/B Testing} 

The distilled AdsGNN$_{t}$ model has already been successfully deployed in a major sponsored search platform and demonstrated significant performance gains. 
Revenue Per Mile (RPM) and Defect Rate are selected as measurements to estimate the revenue gained for every
thousand search requests and the ratio of irrelevant ad impressions, respectively.   
The defected impressions are labeled by human experts.   
C-DSSM model is the original online severing model. 
The online A/B testing results of the AdsGNN$_{t}$ model are summarized in Table \ref{tab:online}.  
Our proposal is employed on both recall and relevance
stage of the ads search. 
As shown in the table, the AdsGNN$_{t} $ model achieves very impressive results because it can significantly increase RPM and reduce advertising defect rates. 
Results demonstrate that our proposal is capable of improving the user experience and driving revenue for the advertisers simultaneously.

\section{Related Work}\label{RelatedWork}
In this section, we will summarize the related works.  
Relevance modeling in sponsored search can be viewed as the sub-domain of information retrieval (IR). 
In terms of matching queries with documents, traditional approaches usually adopt the shallow language representation learning models such as LSA \cite{salakhutdinov2009semantic}, LDA \cite{Blei2012Latent} and Bi-Lingual Topic Models \cite{2011Clickthrough}. 
In recent years, deep learning based approaches have dominated the search area. 
Specifically, the siamese structure is adopted in a range of works \cite{2014Learning, 2014A, 2015Convolutional, 2015Improved, 2017Modeling,li2017ppne}.   
By modeling local contextual information at the word level, Shen et al. \cite{2014Learning} present a series of latent semantic models based on CNN to learn semantic vectors for web searching.
To map source-target document pairs into latent semantic space, Gao et al. \cite{2017Modeling} propose a deep semantic similarity model with special convolutional-pooling structure. 
Besides, several works adopt the interaction-based structure \cite{2015A,  Yin2015ABCNN, Yang2018aNMM}. 
Yang et al. \cite{Yang2018aNMM} propose an attention-based neural matching model that adopts a value-shared weighting scheme to combine matching signals. 
Auto-encoders are used in \cite{salakhutdinov2009semantic}, in which documents are mapped to memory addresses so that semantically similar documents are located at nearby addresses. 
In addition, several works also utilize lexical and semantic matching networks \cite{guo2016deep, mitra2017learning}. By employing a jointly learning framework at the query term level, Guo et al. \cite{guo2016deep} propose a deep relevance matching model for ad-hoc retrieval. With two separated deep neural networks, Mitra et al. \cite{mitra2017learning} propose a document ranking model utilizing local representation and learned distributed representations respectively to match the query and document. Bai et al. \cite{bai2018scalable}  propose an n-gram based embedding of queries and ads to perform the efficient sponsored search. 
Grbovic et al. \cite{grbovic2018real} propose a real-time personalization solution where embeddings of items that user most recently interacted with are combined in an online manner to calculate similarities to items that need to be ranked. 
Huang et al. \cite{huang2020embedding} design a unified embedding framework developed to model semantic embeddings for personalized search.

Apart from the textual information, some related works attempt to incorporate other types of data in sponsored search scenario.  
Yang et al. \cite{yang2019learning} propose to learn the compositional representations by fusing the textual, visual and relational data. 
Zhu et al. \cite{zhu2021textgnn} extend the Twin-Bert model to TextGNN, which aims to incorporate the search log data as complementary. 
Different from existing works \cite{zhu2021textgnn} only focusing on a single aspect of data fusion,  we aim to extensively investigate how to naturally fuse the semantic textual information and the user behavior graph from different perspectives and address several critical challenges in industry scenario.

\begin{table}
	\centering
	\begin{threeparttable}
		\caption{Results of online A/B test.}
		\begin{tabular}{ccc}
			\toprule
			\multicolumn{1}{c}{}&\multicolumn{1}{c}{Relative RPM}&\multicolumn{1}{c}{Relative Defect Rate}\cr
			\midrule
			Relevance phase & +1.13\% &  -1.46\%  \\
			Recall phase & +1.02\% &  -1.17\% \\
			\bottomrule
		\end{tabular}
		\label{tab:online}
	\end{threeparttable}
\end{table} 

\section{Conclusion}
In this paper, we propose a set of AdsGNN models to effectively incorporate the click graph as complementary from three aspects to improve the relevance matching performance. 
Three variations are designed to aggregate the neighborhood information from node-, edge- and token-levels. 
After that, two domain-specific pre-training tasks are proposed to warm up model parameters.
A knowledge-distillation model is further proposed to handle the challenge of long-tail entity inference.   
Empirically, we evaluate the proposed models over a real-life dataset, and the experimental results demonstrate the superiority of our proposal.

\bibliographystyle{ACM-Reference-Format}
\bibliography{ijcai20}

\end{document}